\documentclass[twocolumn,tighten,numberedappendix]{aastex63}
\usepackage{amsmath}
\setcitestyle{notesep={ }}

\graphicspath{{./}{figures/}}

\newcommand\alfven{Alfv\'{e}n~}

\def\simlt{\lower.5ex\hbox{$\; \buildrel < \over \sim \;$}}
\def\simgt{\lower.5ex\hbox{$\; \buildrel > \over \sim \;$}}

\def\beq{\begin{equation}}
\def\eeq{\end{equation}}
\def\ba{\begin{eqnarray}}
\def\ea{\end{eqnarray}}

\def\Eq{Equation}

\def\LA{L_{\rm A}}

\def\M{{\cal M}}
\def\Lsd{L_{\rm sd}}

\def\n{\tilde{n}}
\def\B{{\cal B}}
\def\D{\tilde{\Delta}}

\def\Lf{L_{\rm f}}
\def\Gw{\Gamma_{\rm w}}
\def\Lw{L_{\rm w}}

\def\sigw{\sigma_{\rm w}}

\def\rL{r_{\rm L}}

\def\E{{\cal E}}
\def\N{{\cal N}}
\def\I{{\cal I}}

\def\RLC{R_{\rm LC}}

\def\tobs{t_{\rm obs}}
\def\Ew{{\cal E}_{\rm w}}
\def\EFRB{{\cal E}_{\rm FRB}}

\def\Gsh{\Gamma_{\rm sh}}

\def\RNS{R_{\star}}
\def\BNS{B_{\star}}

\newbox\grsign \setbox\grsign=\hbox{$>$} \newdimen\grdimen \grdimen=\ht\grsign
\newbox\simlessbox \newbox\simgreatbox \newbox\simpropbox
\setbox\simgreatbox=\hbox{\raise.5ex\hbox{$>$}\llap
     {\lower.5ex\hbox{$\sim$}}}\ht1=\grdimen\dp1=0pt
\setbox\simlessbox=\hbox{\raise.5ex\hbox{$<$}\llap
     {\lower.5ex\hbox{$\sim$}}}\ht2=\grdimen\dp2=0pt
\setbox\simpropbox=\hbox{\raise.5ex\hbox{$\propto$}\llap
     {\lower.5ex\hbox{$\sim$}}}\ht2=\grdimen\dp2=0pt
\def\simgt{\mathrel{\copy\simgreatbox}}
\def\simlt{\mathrel{\copy\simlessbox}}

\begin{document}

\title{
Plasmoid ejection by Alfv\'en waves and 
the fast radio bursts from 
SGR 1935+2154}

\author[0000-0002-0108-4774]{Yajie Yuan}
\affiliation{Center for Computational Astrophysics, Flatiron Institute, 162 Fifth Avenue, New York, NY 10010, USA}

\author{
Andrei M. Beloborodov
}
\affil{
Physics Department and Columbia Astrophysics Laboratory, Columbia University, 538  West 120th Street New York, NY 10027}
\affil{
Max Planck Institute for Astrophysics, Karl-Schwarzschild-Str. 1, D-85741, Garching, Germany
}

\author[0000-0002-4738-1168]{Alexander Y. Chen}
\affil{Department of Astrophysical Sciences, Princeton University, Princeton, NJ 08544, USA}

\author{Yuri Levin}
\affiliation{Center for Computational Astrophysics, Flatiron Institute, 162 Fifth Avenue, New York, NY 10010, USA}
\affil{
Physics Department and Columbia Astrophysics Laboratory, Columbia University, 538  West 120th Street New York, NY 10027}
\affil{Department of Physics and Astronomy, Monash University, Clayton VIC 3800, Australia}

\correspondingauthor{Yajie Yuan}
\email{yyuan@flatironinstitute.org}
\correspondingauthor{Andrei M. Beloborodov}
\email{amb@phys.columbia.edu}

\begin{abstract}
    Using numerical simulations we show that low-amplitude Alfv\'en waves from a
    magnetar quake propagate to the outer magnetosphere and convert to
    ``plasmoids'' (closed magnetic loops) which accelerate from the star,
    driving blast waves into the magnetar wind. Quickly after its formation, the
    plasmoid becomes a thin relativistic pancake. It pushes out the
    magnetospheric field lines, and they gradually reconnect behind the pancake,
    generating a variable wind far stronger than the normal spindown wind of the
    magnetar. Repeating ejections drive blast waves in the amplified wind. We
    suggest that these ejections generate the simultaneous X-ray and radio
    bursts detected from SGR~1935+2154. A modest energy budget of the
    magnetospheric perturbation $\sim 10^{40}$~erg is sufficient to produce the
    observed bursts. 
    Our
    simulation predicts a narrow (a few ms) X-ray spike from the magnetosphere,
    arriving almost simultaneously
    with the radio burst emitted far outside the
    magnetosphere. This timing is caused by the extreme relativistic motion of
    the ejecta.
\end{abstract}



\section{Introduction}

Neutron stars with ultrastrong magnetic fields $B\sim 10^{14}$-$10^{16}\,$G
\citep[dubbed ``magnetars'' by][]{1992ApJ...392L...9D} exhibit extreme X-ray
activity \citep[see][for a review]{2017ARA&A..55..261K}. Magnetars were also
suspected as sources of fast radio bursts (FRBs), and the detection of FRBs from
SGR~1935+2154 on 2020 April 28 has established this connection
\citep{2020arXiv200510324T, 2020arXiv200510828B}. SGR~1935+2154 is a magnetar
residing in our galaxy. It has spin period $P\approx 3.2\,$s and magnetic dipole
moment $\mu\sim 2\times 10^{32}$~G~cm$^3$, which corresponds to a surface
magnetic field $B_\star\sim 2\times 10^{14}\,$G.

Scenarios for FRB emission by magnetars can now be put to test. One mechanism is
an ultrarelativistic ejection, which launches a blast wave far beyond the
magnetosphere (\citealt{2014MNRAS.442L...9L}; \citealt{2017ApJ...843L..26B,
  2020ApJ...896..142B}; hereafter B17, B20). This model invokes synchrotron
maser emission by the collisionless shock from the explosion. The shock can
propagate in the magnetar wind of relativistic $e^\pm$ pairs or in a slow
baryonic outflow (B17, B20; \citealt{2019MNRAS.485.4091M,2020MNRAS.494.4627M}).
It has also been proposed that FRBs can come directly from the neutron star
magnetosphere \citep{2002ApJ...580L..65L,2016ApJ...826..226K,
  2017MNRAS.468.2726K, 2020arXiv200505093L,2020arXiv200506736L}. In this
scenario, a concrete mechanism for coherent emission is yet to be worked out.

The most common form of magnetar activity is the hard X-ray bursts with energies
$\E_X\sim 10^{39}-10^{41}$~erg. This nonthermal activity must be generated by
surface motions of the neutron star, which can both slowly twist the
magnetosphere and quickly launch Alfv\'en waves \citep{1989ApJ...343..839B, 1996ApJ...473..322T}. 
Intriguingly, the two radio bursts of milisecond duration detected from SGR~1935+2154 were observed during an X-ray burst, which lasted $\sim 0.5\,$s and had a total energy $\sim 10^{40}$~erg, assuming a distance of
$\sim$ 10 kpc \citep{2020ApJ...898L..29M,2020arXiv200511071L,2020arXiv200511178R}. Each FRB arrived a few ms ahead of a narrow X-ray spike of energy $\E_X\simlt 10^{39}$~erg, exceeding the FRB energy by a factor $>10^3$.
The energy
budget of magnetospheric perturbations generating this activity is likely a few times $10^{40}\,$erg, depending on the efficiency and beaming of the X-ray
emission. This energy is only $\sim 10^{-6}$ of the total magnetospheric energy.

A question arises whether such low-energy events should produce ejecta from the
magnetosphere, which is essential for
the blast-wave FRB mechanism.
In this Letter we investigate the magnetospheric response to a small shear
perturbation of the magnetar surface. We calculate the magnetospheric dynamics
in the framework of force-free electrodynamics (FFE), similar to
\citet{2013ApJ...774...92P}. We use our own finite difference code \emph{Coffee}
(COmputational Force FreE
Electrodynamics)\footnote{\href{https://github.com/fizban007/CoffeeGPU}{https://github.com/fizban007/CoffeeGPU}}
\citep{2020ApJ...893L..38C}. The numerical method is described elsewhere \citep{2020arXiv200711504Y}. The simulation assumes efficient $e^\pm$ creation to
maintain the FFE approximation at low energy costs. This assumption is likely
satisfied by magnetars (B20).

\section{Alfv\'en waves from starquakes}

Sudden excitations of crustal shear oscillations (starquakes) were invoked to
explain X-ray bursts from magnetars \citep{1996ApJ...473..322T}.\footnote{Quakes
  likely involve new forms of crustal response to magnetic stresses, which
  remain to be understood. \citet{2014ApJ...794L..24B} showed that thermoplastic
  instabilities develop in the stressed crust, in particular when the
  ultrastrong $B$ suppresses normal yielding \citep{2012MNRAS.427.1574L}.
  \citet{2017ApJ...841...54T} argued that the crustal yielding can develop
  quickly, on a millisecond timescale.}
Their energy could reach
$E_Q^{\max}\sim V\mu_{\rm sh} s^2/2\sim 10^{44}\,V_{16}(s/0.1)^2\,$erg, where
$\mu_{\rm sh}\sim 10^{30}\,$erg/cm$^3$ is the shear modulus of the deep crust,
$s<0.1$ is the elastic strain, and $V$ is the stressed volume (a fraction of the
deep crust volume $\sim 10^{17}\,$cm$^3$).

After its trigger stops, the quake duration $t_Q$ is limited by the magnetic
coupling of the crust to the liquid core
\citep{2006MNRAS.368L..35L,2020ApJ...897..173B}. The coefficient of shear wave
transmission to the core ${\cal T}_c\sim 0.1 B_{14}^{1/2}$ is controlled by the
poloidal magnetic field $B$. The waves will bounce at least
$\sim {\cal T}_c^{-1}$ times, spreading through the crust. They propagate with
speed $\sim 10^8$~cm/s, and each quake should last tens of milliseconds, with
wave frequencies $\nu\simgt 1\,$kHz. A series of quakes can make the activity
much longer.

The quake launches Alfv\'en waves into the magnetosphere
\citep{1989ApJ...343..839B, 1996ApJ...473..322T, 2020arXiv200108658B}. Waves are
launched on the magnetospheric field lines whose length exceeds the wavelength
$\lambda= c/\nu\approx 3\times 10^7\,\nu_{\rm kHz}^{-1}$~cm, which is much
greater than the star's radius $\RNS\approx 10^6\,$cm.

Far from the star, the magnetic field lines are approximately dipole and reach
their maximum radii $R$ at the magnetic equator. A field line extending to
radius $R\gg \RNS$ has the approximate length $\sim 3R$. Thus, the waves are
launched on field lines extending to
\begin{equation}
   R> 10^7\,\nu_{\rm kHz}^{-1}\,{\rm cm}.
\end{equation}
The magnetic flux reaching the sphere of radius $R$ is $\Psi=2\pi\mu/R$ (where
$\mu$ is the magnetic dipole moment). The footprint area of flux $\Psi$ on the
star's surface is $A\sim \Psi/\BNS$, where $\BNS$ is the surface magnetic field.
In particular, for an approximately dipole magnetosphere,
$\mu\sim B_\star \RNS^3$ and $A\sim 2\pi \RNS^3/R$.

The Alfv\'en waves emitted along the extended field lines over the quake duration $t_Q$ carry a small fraction $f$ of the quake energy
\citep{2020ApJ...897..173B}, with power
\begin{equation}
\label{eq:LA}
\LA\sim 10^{42} \left(\frac{f}{0.01}\right)
\left(\frac{E_Q}{10^{42}\,{\rm erg}}\right)
\left(\frac{t_Q}{10\,{\rm ms}}\right)^{-1} \frac{\rm erg}{\rm s}.
\end{equation}

The relative amplitude of the wave $\delta B/B$ is small near the star, $\delta B_\star/B_\star\ll 1$, and grows with radius $r$,
\begin{equation}
\label{eq:growth}
   \frac{\delta B}{B}=\frac{\delta B_\star}{B_\star} \left(\frac{r}{R_\star}\right)^{3/2}. 
\end{equation}

\section{Plasmoid ejection}

\begin{figure*}[t]
    \centering
    \includegraphics[width=\textwidth]{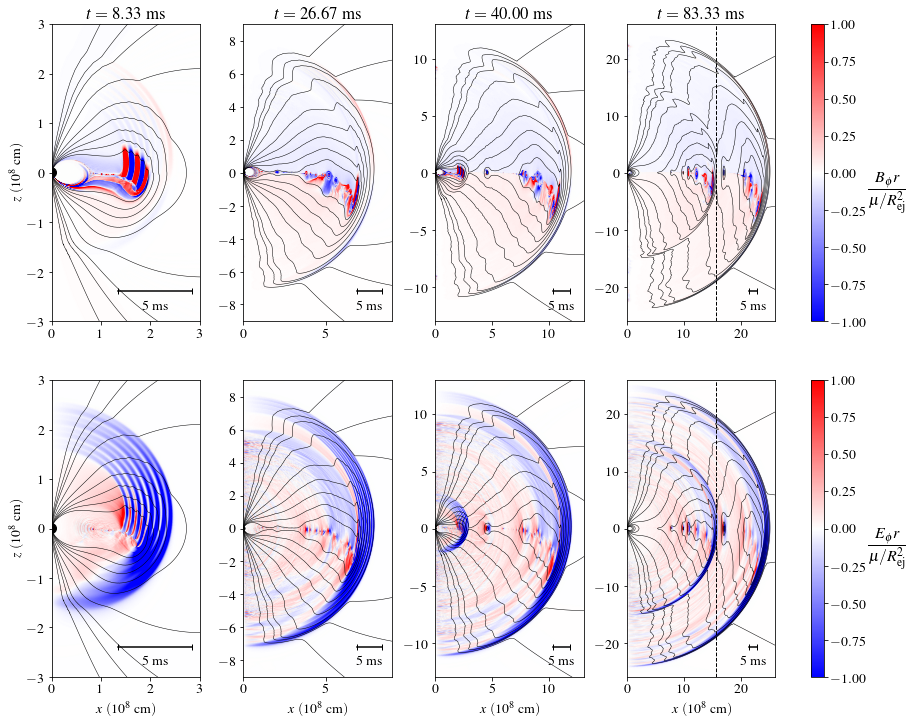}
    \caption{
Four snapshots of the simulation.
Thin black curves are the 
poloidal magnetic field lines.
Color shows $rB_\phi$ (top) and $rE_\phi$ (bottom).
Lengths are normalized to the ejection radius $R_{\rm ej}=10^8\,$cm, and fields are normalized to $B_0=\mu/R_{\rm ej}^3$, where $\mu$ is the magnetic dipole moment of the star. 
The vertical dashed line indicates
the light cylinder $\RLC=c/\Omega$.
}
    \label{fig:BE}
\end{figure*}

\begin{figure*}[t]
    \centering
      \includegraphics[width=0.8\textwidth]{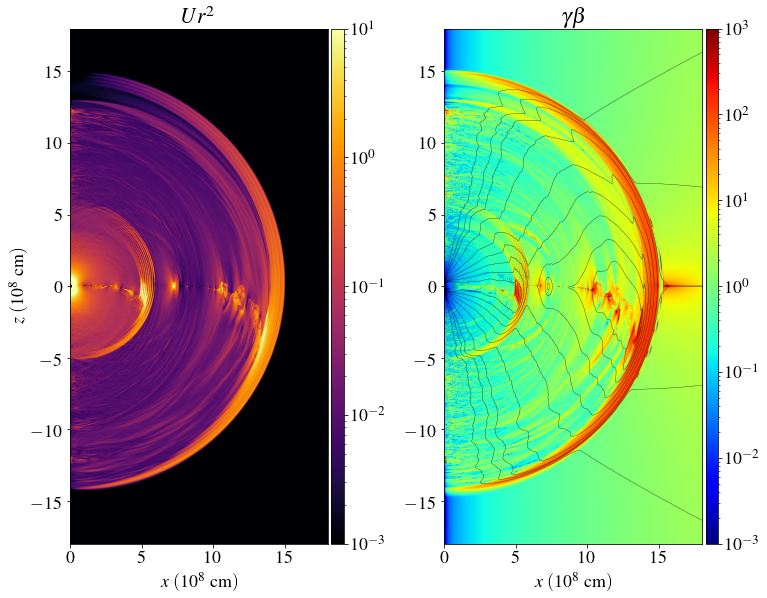}
    \caption{Left: energy density
$U=(B^2+E^2)/8\pi$.
Right:
$\mathbf{E}\times\mathbf{B}$ drift 4-velocity.
The snapshot is taken at $t=50\,$ms. Units are the same as in Figure~1.
}
    \label{fig:Uvj}
\end{figure*}

\begin{figure*}
    \centering
    \includegraphics[width=\textwidth]{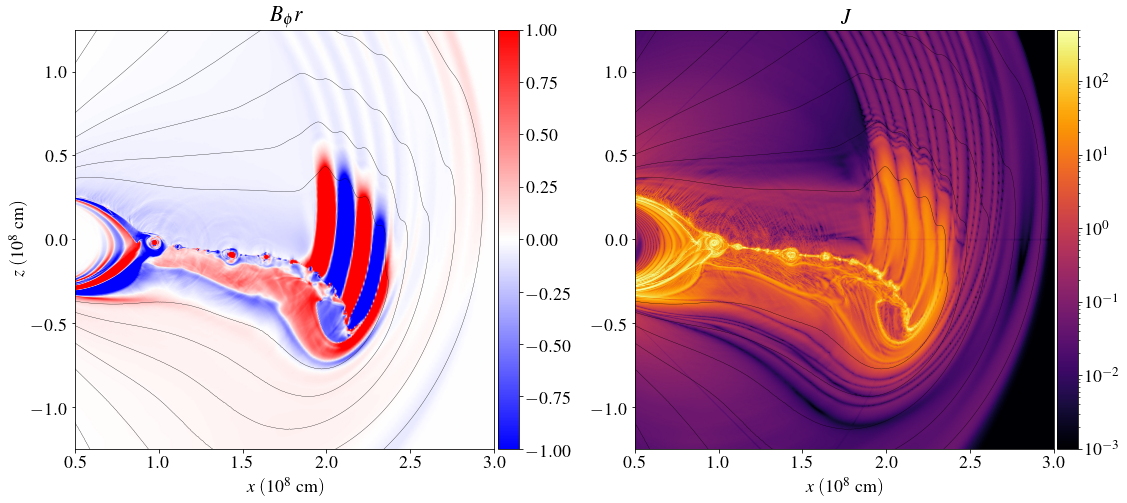}
    \caption{Zoom-in view of the ejecta at $t=10\,$ms. 
Left: toroidal field $B_\phi$. Right: magnitude of current $J$. Units are the same as in Figure~1.
}
    \label{fig:BJzoom}
\end{figure*}

The wave growth with $r$ leads to $\delta B>B$ on sufficiently extended field
lines, $R/\RNS>(\delta \BNS/\BNS)^{-2/3}$. Then, as demonstrated below, the wave
generates a closed magnetic island (``plasmoid'') and ejects it from the
magnetosphere. This occurs far from the star, where the magnetospheric energy
$\sim R^3B^2$ is comparable to the wave energy $\Ew$. The ejecta energy is
\begin{equation}
\E_p\sim \E_w\sim R^3 B^2\sim\frac{\mu^2}{R^3} = 10^{40}\,\mu_{32}^2\,R_8^{-3}\,{\rm erg}.
\end{equation}

Our simulation is axisymmetric and starts with a steady-state, force-free dipolar magnetosphere attached
to a rotating star (with aligned magnetic and rotational axes). At $t=0$, a
sinusoidal Alfv\'en wave is injected by shearing the footprints of the closed
magnetic field lines that extend to $R\sim 10^8$~cm. We injected a packet of
four wavelengths $\lambda=2.5\times 10^7$cm to demonstrate one plasmoid
formation, and then (30~ms later) another packet, to illustrate multiple
plasmoid ejections. The energy of each packet is
${\cal E}_w\approx 9\times 10^{40}$~erg. The initial wave amplitude is small,
$\delta B_\star/B_\star\approx 0.004$, and its propagation for a while follows
\Eq~(\ref{eq:growth}).\footnote{Our simulation verified the law
  $\delta B/B=(\delta \BNS/\BNS)(r/R_\star)^{3/2}$ during the linear phase
  $\delta B\ll B$, and then we removed the trivial linear evolution phase by
  moving the inner boundary of the simulation box from $\RNS$ to
  $R_{\rm inj}\approx 10 \RNS$ while correspondingly increasing the injected
  amplitude by $(R_{\rm inj}/\RNS)^{3/2}\approx 32$, from
  $\delta B_\star/B_\star\approx 0.004$ at $\RNS$ to $\delta B/B\approx 0.13$ at
  $R_{\rm inj}$.} Interesting nonlinear evolution occurs at $r\sim 10^8\,$cm
where $\delta B$ exceeds $B$ (Figure~\ref{fig:BE}).


We observe that at the radius $R_{\rm ej}\approx 10^8\,$cm the Alfv\'en wave
``breaks'' and forms a plasmoid with energy $\E_p\sim 0.6\,\E_w$. The plasmoid
immediately accelerates away, pushing its way through the outer magnetosphere.
The ambient magnetic energy decreases with radius as $r^3B^2\propto r^{-3}$ and
quickly becomes negligible compared with $\E_p$, so the plasmoid continues to
expand freely, unaffected by the background. Its transverse size scales linearly
with $r$ while its radial thickness remains approximately constant,
$\Delta\sim 10^8\,$cm, so the plasmoid becomes a thin pancake. Its Lorentz
factor quickly grows to $\gamma> 10^2$ (Figure~\ref{fig:Uvj}). The pancake
occupies an extended solid angle around the
magnetic equator.
Most of its energy
is contained in an angular range $\Delta\theta\sim 0.4$.

The pancake structure reflects the initial shape of the Alfv\'en wave at
$r\ll R_{\rm ej}$ and the process of its breakout at $R_{\rm ej}$. The \alfven
wave carries the perturbations $B_{\phi}$ and $E_{\theta}$, which are supported
by an electric current $J$
along the background dipole magnetic field. As the wave
packet breaks away, these field components and the current are advected with the
plasmoid. In addition, a strong electromagnetic wave of $E_\phi$ and
$\delta B_{\theta}$ (with $J=0$)
is launched ahead of the
current-carrying plasmoid.

This ejecta forces the magnetosphere to open up, creating a Y-shaped current
sheet separating the opposite magnetic fluxes in the two hemispheres
(Figure~\ref{fig:BJzoom}). The current sheet is unstable to reconnection, and
the opposite magnetic fluxes combed-out by the pancake gradually snap back
behind it, ejecting numerous small-scale plasmoids of various sizes and
energies. A similar plasmoid chain formed in the flare simulations of
\citet{2013ApJ...774...92P}. The equatorial current sheet extends to the
southern end of the pancake; reconnection also occurs there (Figure
\ref{fig:BJzoom}).

\begin{figure*}
    \centering
    \includegraphics[width=\textwidth]{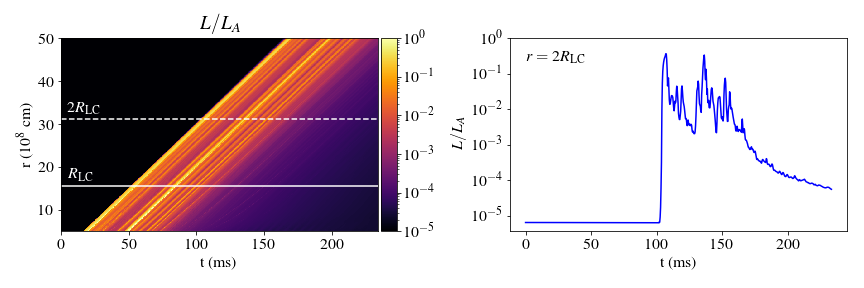}
    \caption{Left: net Poynting flux $L(t,r)$ (through spheres of different radii $r$)
normalized to the \alfven wave luminosity $\LA$. 
Right: $L(t)$ at $r=2\RLC$ (along the dashed line in the left panel).}
    \label{fig:poynting}
\end{figure*}

Reconnection at the ejection site continues to generate a variable outflow until
the second Alfv\'en wave packet arrives at $R_{\rm ej}$, and then the second
pancake is ejected (Figures~\ref{fig:BE}--\ref{fig:Uvj}). The entire region
between the two pancakes is filled with the variable wind generated by magnetic
reconnection. This wind forms the ambient medium encountered by the second
pancake. The wind power $\Lw(t)$ is much lower than the pancake power
$\sim \E_p c/R_{\rm ej}\sim \LA$ and much greater than the spindown power of the
magnetar $L_0\approx \mu^2\Omega^4/c^3$ (Figure~\ref{fig:poynting}).

After crossing the light cylinder $\RLC=c/\Omega$, the leading pancake begins to
sweep the torodial spindown wind of the magnetar. The second pancake continues
to sweep the variable flow between the two pancakes.

In our simulation, the light cylinder is located at $\RLC=1.5\times 10^{9}$~cm,
which is 10 times smaller than $\RLC$ in SGR~1935+2154, i.e. our star rotates 10
times faster. This choice was made to accommodate $\RLC$ in the computational
domain\footnote{The simulation is axisymmetric and performed in spherical
  coordinates $r,\theta,\phi$. We use a $8192 \times 4096$ grid with uniform
  spacing in $\ln r$ and $\theta$, covering the region
  $10^7<r<5\times 10^{9}$~cm.} well inside of its outer boundary
$R_{\rm out}=5\times 10^{9}$~cm.
However, the pancakes and the wind between them formed at $r\ll\RLC$ unaffected
by the magnetar rotation, so their parameters need no rescaling.

The simulation shows that about 60\% of the initially injected wave energy is
carried away by the ejecta, and 10-20\% is dissipated. The dissipation occurs in
the current sheets that form during the ejection process
(Figure~\ref{fig:BJzoom}). It is captured only approximately in FFE simulations,
via three numerical channels. (1) To satisfy the FFE conditions, at every
timestep we remove any $\mathbf{E}\cdot\mathbf{B}$ by resetting
$\mathbf{E}\to\mathbf{E}-(\mathbf{E}\cdot\mathbf{B})\mathbf{B}/B^2$. (2)
Whenever $E>B$ happens, we reset $\mathbf{E}$ to $(B/E)\mathbf{E}$. (3) We apply
the standard
suppression of 
high frequency noise 
\citep{kreiss_methods_1973}. 
The resulting numerical dissipation occurs mainly in the thin current sheets, and serves as a proxy of 
physical dissipation. 

\begin{figure*}[t]
    \centering
    \includegraphics[width=\textwidth]{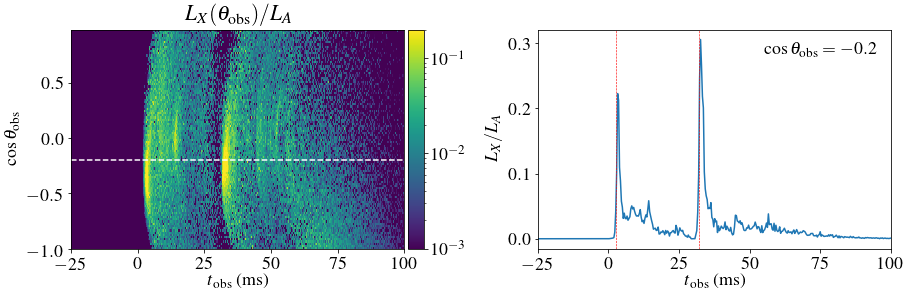}
    \caption{Left: distribution of apparent X-ray luminosity $L_X(\theta_{\rm obs})$ over polar directions $\theta_{\rm obs}$ and arrival times $\tobs$. Right: X-ray light curve seen by an observer at $\cos\theta_{\rm obs}=-0.2$ (white dashed line in the left panel). The vertical red dashed lines mark the arrival times of the blast-wave emission at $r\simgt 10^{13}\,$cm.}
    \label{fig:lightcurve}
\end{figure*}

Assuming that the dissipated energy is emitted isotropically in the local plasma
rest frame, we have calculated the bolometric light curve of this emission
(Figure~\ref{fig:lightcurve}). The plasma moves with velocity
$\mathbf{v}=\mathbf{E}\times\mathbf{B}/B^2$, and part of the dissipation occurs
where the ejecta have already accelerated, which leads to strong Doppler
boosting. In particular, substantial emission comes from the current sheet at
the southern tail of the pancake, which develops a high Lorentz factor $\gamma$
(Figure~\ref{fig:BJzoom}), and there is a similar current sheet in the second
pancake. This results in the two strong spikes in the light
curve.\footnote{Figure~\ref{fig:lightcurve} includes dissipation at
  $r<6 R_{\rm ej}$, as the declining resolution $\delta r\propto r$ complicates
  dissipation measurements at $r\simgt 10R_{\rm ej}$. Including $r>6 R_{\rm ej}$
  would make the two spikes higher.} A significant part of this emission should
appear in the X-ray band, and thus, two X-ray spikes are predicted by the
simulation. 
The centroid of the X-ray spike is delayed by a few ms relative to the blast wave emission (Figure~5).

\section{FRB emission from the explosion}

\subsection{Shock maser emission}

An observer at polar angle $\theta$ will see the pancake with the apparent energy
\begin{equation}
  \E(\theta)= 4\pi \frac{d\E_p}{d\Omega}=b(\theta)\,\E_p.
\end{equation}
In our simulation, the beaming factor $b(\theta)$ is found to vary from $b\ll 1$
to $b\sim 5$ for favorable lines of sight. The isotropic equivalent of the
explosion power $\Lf(\theta)$ is related to the original Afv\'en wave power
$\LA$ as
\begin{equation}
   \Lf(\theta)\sim b(\theta) \LA. 
\end{equation}
It will drive an ultrarelativistic blast wave in the wind.

Calculations of 
maser radio emission from the blast wave are based on kinetic simulations of
collisionless shocks 
\citep[e.g.,][]{2019MNRAS.485.3816P}
and give the following (see
B20). If the GHz waves are emitted before the blast wave begins to decelerate,
the radio burst has the apparent energy
\begin{equation}
\label{eq:EFRB}
  \EFRB\sim 
  10^{-3} \, \Lf\, \delta \tobs \sim 10^{36}\,L_{\rm f,42} \left(\frac{\delta\tobs}{1\,\rm ms}\right)  {\rm erg}.
\end{equation}
The burst duration $\delta\tobs$ is
\begin{equation}
\delta\tobs\sim \frac{r}{c\Gsh^2}, \qquad \Gsh
=2\sigw^{1/2}\Gw\left(\frac{\Lf}{\Lw}\right)^{1/4}.
\end{equation}
Here $\Gsh$ is the shock Lorentz factor; $\Lw$, $\sigw$, $\Gw$ are the power,
magnetization parameter, Lorentz factor of the upstream wind swept by the shock.

Consider the wind behind pancake~1. It serves as the external medium for
pancake~2 (Figure~\ref{fig:Uvj}). The wind power and speed profiles grow outward, reaching
maximum at pancake 1. The blast wave from pancake~2 will chase the wind layers,
so that at radius $r$ it picks up layers with $\Gw^2\sim r/cT$, where
$T\sim 30\,$ms is the time separating the two pancakes. This gives
\begin{equation}
\label{eq:tobs}
   \delta\tobs\sim\frac{T}{4\sigw}\left(\frac{\Lw}{\Lf}\right)^{1/2}
   \sim \frac{1\,\rm ms}{\sigw} \left(\frac{T}{30\,\rm ms}\right) 
   \left(\frac{L_{\rm w,40}}{L_{\rm f,42}}\right)^{1/2}.
\end{equation}
The shock emits in the GHz band at radius
\begin{equation}
   R_{\rm GHz}\sim 10^{13}\,L_{\rm f,42}^{1/4} L_{\rm w,40}^{1/4} {\rm~ cm}.
\end{equation}
$\Lf$ depends on the viewing angle $\theta$ and decreases outside
favorable $\theta$. For instance, $\Lf\sim 10^{41}\,$erg/s gives
$\EFRB\sim 10^{35\,}$erg and $\delta \tobs\sim 3\sigw^{-1}\,$ms, suitable for
the FRBs from SGR~1935+2154.

Multiple ejections can result in multiple blast waves in the enhanced wind, and
thus produce multiple FRBs. The small $\delta\tobs$ implies that their arrival
times closely track the
pancake
ejection time, and so the FRBs arrive
simultaneously with the X-rays generated by the ejections
(Figure~\ref{fig:lightcurve}).

\subsection{Magnetosonic waves carried by the pancake}

In our simulation we observed formation of current sheets and waves of various
scales during the 
pancake
ejection. High-frequency fluctuations are expected
from magnetic reconnection in the current sheets: reconnection forms a
self-similar plasmoid chain extending to a microscopic kinetic scale comparable
to the particle Larmor radius $r_{\rm L}$ \citep[e.g.][]{2010PhRvL.105w5002U}.
The chain is observed in both kinetic \citep{2014ApJ...783L..21S} and FFE
\citep{2013ApJ...774...92P} simulations. FFE simulations truncate the chain on
the grid scale as they do not follow kinetic processes.

Small-scale plasmoid mergers in the chain
generate fast magnetosonic waves with frequencies
up to $\nu\sim c/r_{\rm L}$ and energy flux $F\simlt 10^{-4}cB^2/4\pi$, where
$B$ is the reconnecting field \citep{2019MNRAS.483.1731L,2019ApJ...876L...6P}.
The high-frequency waves can carry up to $10^{-4}$ of the energy released by
reconnection, comparable to $10^{-4}\E_p$. The 
pancake
ejection in our
simulation occurs at radii $r<10^9~\rm{cm}$ where $B>10^7\,$G. The corresponding $\rL$
implies that reconnection generates GHz waves, although our simulation can only
resolve waves from much larger plasmoids (Figure~\ref{fig:BE}).
 
In FFE, fast magnetosonic modes are indistinguishable from vacuum
electromagnetic waves. In reality, the generated waves live in the plasma of a
finite inertia; they are advected with the pancakes. At large radii (far outside
the magnetosphere) the waves may escape as radio waves
\citep{2020ApJ...897....1L} if they are not damped before reaching the escape
radius. If only waves carried by the ultrarelativistic pancakes survive, the
radio bursts would have durations comparable to the pancake thickness
$\Delta/c\sim 3\,$ms. This would be consistent with FRBs from SGR~1935-2154.

\section{Discussion}

Our results demonstrate that flares can occur in the outer magnetosphere and
eject plasmoids similarly to the rare, extremely energetic (``giant'') flares in
the inner magnetosphere. The outer magnetosphere is overtwisted by the outgoing
and growing Alfv\'en wave  \citep[while simulations of giant flares invoked
quasi-static twisting, e.g.,][]{2013ApJ...774...92P,2019MNRAS.484L.124C}.
The resulting explosion power $\Lf$ is comparable to the power of the Alfv\'en
wave launched by the quake, $\LA$. Plasmoid ejection by field lines extending to
a given radius $R$ has an energy threshold
$\E_{\rm thr}\sim \mu^2/R^3=10^{40}\,\mu_{32}^2\,R_8^{-3}\,$erg. During the
quake activity of a magnetar, the ejections may occur intermittently,
depending on the amplitude and location of the quake trigger.

We conclude that a broad range of magnetic ejections with
$\Lf\sim 10^{41}$-$10^{47}\,$erg/s can occur in magnetars. The picture of blast
waves from magnetic flares (B17, B20) then implies a broad range of FRB energies
scaling with $\Lf$ (Equation~\ref{eq:EFRB}) and produced by the same mechanism.
It includes the repeating superstrong FRBs from young, hyper-active magnetars in
distant galaxies and the weak bursts from the local, older magnetar
SGR~1935+2154.

Our simulation shows a huge enhancement of the magnetar wind,
$\Lw>10^{-3}\LA\gg \Lsd$, during the bursting period. Hence, blast waves can
propagate in winds much denser than the normal spindown wind. The first blast
wave, from pancake~1, may not produce a bright FRB because the wind ahead of it
is weak, but the strong wind in the wake of pancake~1 leads to efficient FRB
production by the blast wave from pancake 2. The wind is variable and modulates
the temporal and spectral structure of the shock maser emission (see Section~6.6
in B20). If the pancake tails are indeed the sites of radio emission, three
close ejections may be needed to produce two FRBs detected in SGR~1935+2154.
Such conditions, along with a favorable line of sight, are rare. This may
explain why only a small fraction of magnetar X-ray bursts are accompanied by
FRBs.

Current sheets formed in the wave breakout process are sources of magnetosonic
waves, which likely extend to GHz frequencies. Similar waves from magnetic
flares are discussed by \citet{2020ApJ...893L...6M} in the context of binary
pulsars. If these waves are not damped during subsequent expansion to a much
larger radius where they could escape as radio waves, they could provide another
source of radio emission. Then more frequent FRBs may be expected from
magnetars, as such waves likely accompany every pancake ejection.

Another interesting result of our simulation is the X-ray spike simultaneous
with the blast wave emission. The spike originated from dissipation during the
plasmoid ejection, which took place around the magnetic equator and triggered
magnetic reconnection. We anticipate that the Alfv\'en-wave-driven flares can
eject plasmoids at different polar angles, with different dissipation
rates, and we do not expect a universal energy ratio of the radio and X-ray
spikes, $\E_{\rm radio}/\E_X$.

Our simulation is just one example illustrating magnetic flares in the outer
magnetosphere, which was not fine-tuned to a specific observation. In
particular, our flare energy was probably somewhat higher than needed for the
April 28 event in SGR~1935+2154, depending on the viewing angle. Magnetar flares
are diverse in energy, and their details should depend on the initial shape and
location of Alfv\'en wave emission. We leave their systematic study to future
work.

Future work can advance the model in a few ways. (1) While our FFE simulations
reliably demonstrate the formation of dissipative current sheets, accurate
calculations of dissipation will require kinetic modeling. (2) Our axisymmetric
simulation shows reconnection of poloidal magnetic field lines, and tracing
reconnection of $B_\phi $ will require full 3D simulations. (3) 3D simulations
could model non-axisymmetric Alfv\'en waves and ejecta interacting with striped
winds of inclined rotators. \cite{2020ApJ...897....1L} proposed that this
interaction could generate FRBs with frequency $\nu_{\rm FRB}\propto \Lf^{5/8}$.
When applied to explosions in SGR~1935+2154, it gives
$\nu_{\rm FRB}\simlt 1\,$MHz, however his model is potentially viable for
brighter FRBs.

The possibility of blast waves from SGR~1935+2154 is also discussed by
\citet{2020ApJ...899L..27M}, without specifying the mechanism of the low-energy
explosion. They consider a blast wave hitting a slowly expanding baryonic cloud
at $r\sim 10^{11}\,$cm, and find that it would generate radio and X-ray emission
with $\E_{\rm radio}/\E_X\sim 10^{-5}$.

\bigskip
We thank the referee for helpful comments on the manuscript. Y.Y. is supported by a Flatiron Research Fellowship at the Flatiron Institute, Simons Foundation. A.M.B. is supported by NASA grant NNX\,17AK37G, NSF grant AST\,2009453, Simons Foundation grant \#446228, and the Humboldt Foundation. A.C. is supported by NASA grant 80NSSC18K1099. Y.L. is supported by NSF grant AST\,2009453.

\bibliography{ref}{}
\bibliographystyle{aasjournal}

\end{document}